\documentclass[journal]{IEEEtran}%

\usepackage{graphicx}

\usepackage{amssymb}
\usepackage{amsmath,amsfonts,amssymb}
\usepackage{verbatim}

\usepackage[bookmarks=false]{}

\IEEEoverridecommandlockouts

\begin{document}
%
\title{Selective Multipath Interference Canceller with Linear Equalization for DS-UWB Systems with Low Spreading Factor}

\author{\IEEEauthorblockN{Chunhua Geng, Yukui Pei, and Ning Ge~\IEEEmembership{Member,~IEEE}}
\thanks{This paper was presented in part at the {\it IEEE International Conference on Ultra-Wideband} ({\it ICUWB}), Vancouver, Canada, 2009.
The authors are with the State Key Laboratory on Microwave and
Digital Communications, Tsinghua National Laboratory for Information
Science and Technology, Tsinghua University. (email:
gengch07@mails.tsinghua.edu.cn, peiyk@tsinghua.edu.cn,
gening@tsinghua.edu.cn) }}




\maketitle

\begin{abstract}

In high rate DS-UWB systems with low spreading factor, the selective
multipath interference canceller with linear equalization (SMPIC-LE)
is developed to alleviate severe multipath interferences induced by
the poor orthogonality of spreading codes. The SMPIC iteratively
mitigates the strongest inter-path interference, inter-chip
interference and inter-symbol interference, while the former two are
unresolvable in conventional RAKE-decision feedback equalizer (DFE)
receivers. The numerical results and complexity analysis demonstrate
that SMPIC-LE with proper parameters provides an attractive overall
advantage in performance and computational complexity compared with
RAKE-DFE. In addition, it approaches the matched filter bound well
as the RAKE finger in SMPIC increases.
\end{abstract}


\section{Introduction}
Ultra-wideband (UWB) is a promising technology for wireless high
rate and short range communications \cite{UWB overview}.
Direct-sequence spreading based UWB (DS-UWB) and multiband
orthogonal frequency-division multiplexing UWB (MB-OFDM UWB) are two
main physical layer standards for high data rate wireless personal
area networks (WPAN) \cite{DS-UWB proposal}, \cite{OFDM-UWB
standard}. Because of the fine properties of coherent processing of
the occupied bandwidth and the widest contiguous bandwidth, DS-UWB
has received considerable attention from both academia and industry
\cite{DS advantage}, \cite{DSSS UWB}.

For high data rate DS-UWB systems supporting transmission rate
ranging from several Megabit per second to more than one Gigabit per
second, most recent research on the receiver design focuses on the
RAKE reception with symbol level decision feedback equalizer (DFE)
\cite{EQ1}-\cite{EQ3}. In practical high rate DS-UWB systems,
limited by state-of-the-art ADC technology, the spreading factor
(SF) cannot be large enough to maintain the ideal orthogonality
between spreading codes \cite{ADC}. Therefore, the conventional
RAKE-DFE receiver would suffer significant performance loss due to
severe inter-path interference (IPI), inter-chip interference (ICI)
and inter-symbol interference (ISI) \cite{Interference
ICC}-\cite{Low-SF}. The former two kinds of interference can not be
mitigated by the RAKE-DFE receiver effectively. Furthermore, in
order to combat the severe ISI induced by long channel delay spread,
the DFE tap number has to be quite large. The demanding
computational complexity of DFE always exceeds that of the RAKE
receiver significantly by far and becomes a heavy burden for system
design.

To resolve the above problems, the selective multipath interference
canceller (SMPIC) with symbol level linear equalization (LE) is
proposed in this paper for practical high rate DS-UWB systems with
low SF. The SMPIC is capable of mitigating the IPI, ICI and ISI by
reconstructing and subtracting the selected strongest multipath
interferences from the received signal in an iterative way. Then the
symbol level LE is concatenated to alleviate the remaining ISI. In
addition, to validate the effectiveness of the SMPIC-LE receiver, we
derive the matched filter bound (MFB), which takes into account such
practical constrains as the sampling rate and the RAKE diversity
order, i.e. the finger number. Simulation results and complexity
analysis show that the proposed SMPIC-LE can achieve similar or even
better performance with much lower computational complexity compared
with the conventional RAKE-DFE receiver in various realistic UWB
channels. Moreover, as the RAKE diversity order increases, the
performance of SMPIC-LE receivers can approach the derived MFB well.

The remainder of this paper is organized as follows. The DS-UWB
system model with the proposed SMPIC-LE receiver is introduced in
Section II. In Section III, the computational complexity and
performance of the SMPIC-LE receiver are analyzed, and the MFB is also
derived. In Section IV, the corroborating simulation results are
presented. Section V summarizes the whole paper.

\section{System Model}
\begin{figure*}[t]
\begin{center}
  \includegraphics[width = 12 cm]{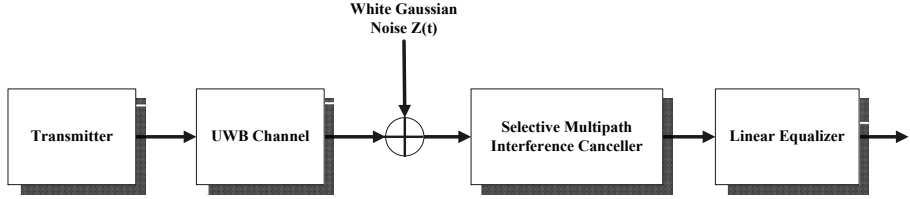}
  \caption{Diagram of the DS-UWB System model with the SMPIC-LE receiver}
  \label{prop_sys}
\end{center}
\end{figure*}

In this paper, the IEEE802.15.3a UWB indoor channel model is
employed \cite{Channel}. The equivalent complex-valued baseband
model with the proposed SMPIC-LE receiver is shown in
Fig.\ref{prop_sys}.

\subsection{Transmitter}
In this paper, we only focus on binary phase-shift keying (BPSK)
modulation, which is the mandatory transmission mode for DS-UWB
systems. At the transmitter, the random source symbol is spread and
modulated with chip pulse $g_T(t)$. For each symbol, the pulse shape
is defined as
\begin{equation}
\label{symbol pulse} g(t)=\sum\limits_{n=0}^{N-1} c[n]g_T(t-nT_c)
\end{equation}
where $c[n]$ denotes the $n$-th chip of the spreading code of length
$N$, and $T_c$ is the chip duration. Assume $M$ symbols are
contained in each frame, and each transmitted frame can be written
as
\begin{equation}
\label{frame} s(t)=\sum\limits_{m=0}^{M-1} b[m]g(t-mT_b)
\end{equation}
where $b[m]\in\{-1,+1\}$ represents the $m$-th symbol of each frame,
and $T_b = NT_c$ is the symbol interval.

\subsection{UWB channels}
In order to compare standardization proposals for high data rate
WPANs, IEEE802.15.3a task group developed a standard channel model
for UWB systems \cite{Channel}. This model is based on the
Saleh-Valenzuela model \cite{S-V} with some modification to account
for the properties of realistic UWB channels. In this model,
multipath arrivals are grouped into two categories: cluster arrivals
and ray arrivals within each cluster. The channel impulse response
is defined as:
\begin{equation}
\label{channel}
h(t)=X\sum\limits_{l=0}^{L-1}\sum\limits_{k=0}^{K-1}\alpha_{k,l}\delta(t-T_l-\tau_{k,l})
\end{equation}
where $X$ represents the log-normal shadowing, $\alpha_{k,l}$ is the
multipath gain coefficient, $T_l$ is the delay of $l$-th cluster and
$\tau_{k,l}$ is the delay of the $k$-th multipath component relative
to the $l$-th cluster arrival time ($T_l$). By definition, we have
$\tau_{0,l} = 0$ for $l\in\{0,1,...,L-1\}$.

\subsection{SMPIC-LE Receiver}

In the DS-UWB system, the SMPIC-LE receiver is developed to
alleviate the severe multipath interferences induced by the poor
orthogonality of spreading codes. In the first stage, the SMPIC is
employed to specifically mitigate the strongest multipath
interference components, including IPI, ICI and ISI. Then, a
conventional symbol level LE with small tap number is concatenated
to combat the residual ISI. In the sequel of this section, we mainly
focus on the proposed SMPIC scheme.

\begin{figure}[t]
\begin{center}
  \includegraphics[width= 8.5 cm]{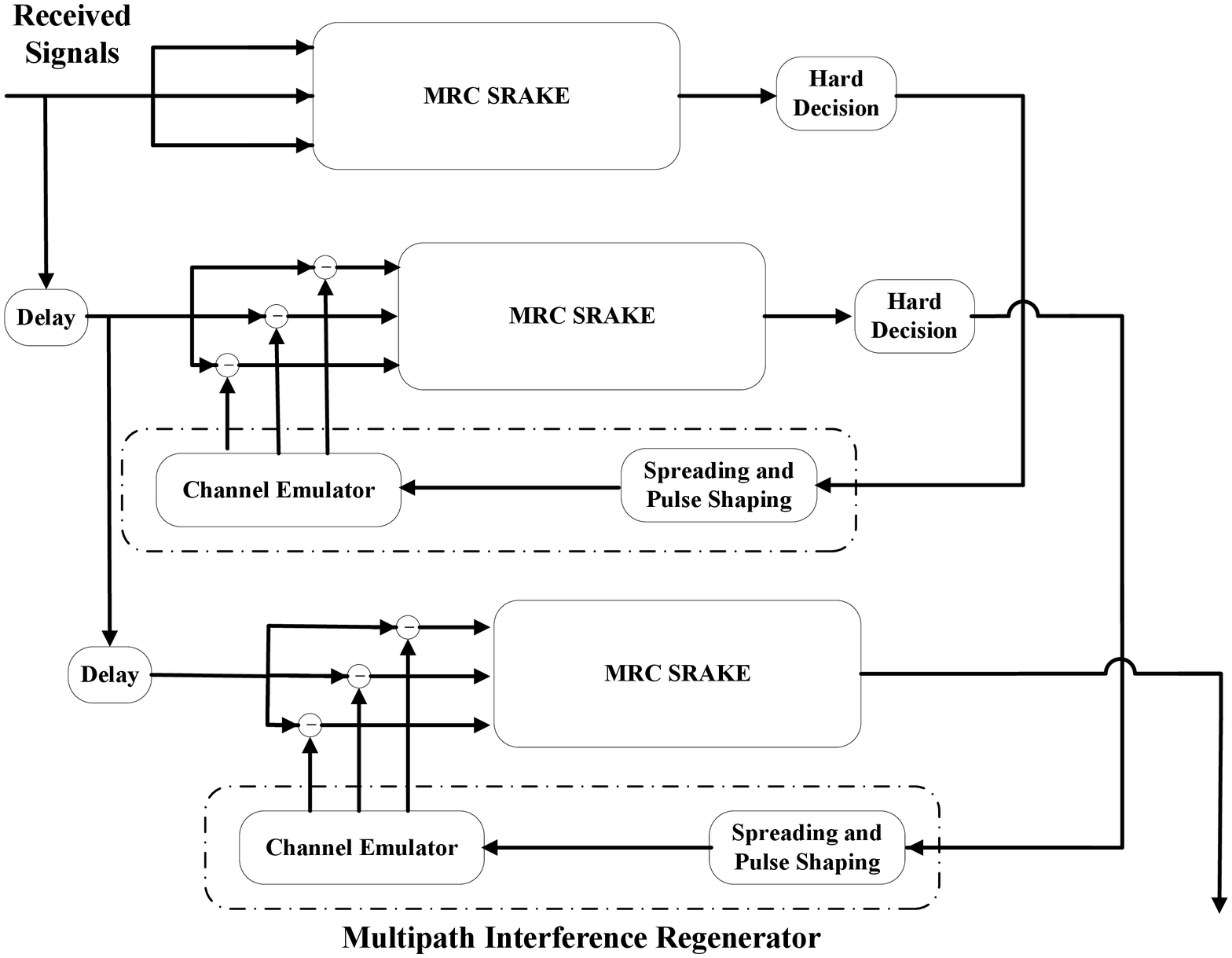}
  \caption{Block diagram of SMPIC (the iteration times $p$ are 2 in this diagram)}
  \label{smpic}
\end{center}
\end{figure}

The structure of SMPIC is presented in Fig.\ref{smpic}. The SMPIC
works in an iterative manner. Its purpose is to eliminate the
interference induced by multipath delay at each RAKE finger. Similar
with selective-RAKE (SRAKE) \cite{S-RAKE}, the SMPIC selects the
instantaneously strongest $J$ multipath components and combines them
together at first. Then the interference is estimated and subtracted
from the received data at each RAKE finger to get more precise input
signals for the RAKE reception in the next iteration. In order to
reduce the complexity, the interference is reconstituted by using
the hard decision of the SRAKE output. Through this iterative
process, the precision of the correlation result in each RAKE finger
is improved, so is the output of the receiver.

In DS-UWB systems, the received signal of each frame is given by
\begin{equation}
\label{received data}
\begin{aligned}
r(t)=
&\sum\limits_{l=0}^{L-1}\sum\limits_{k=0}^{K-1}\sum\limits_{m=0}^{M-1}\sum\limits_{n=0}^{N-1}
a_{k,j}b[m]c[n]\\
&g_T(t-T_l-\tau_{k,j}-nT_c-mT_b)+z(t)
\end{aligned}
\end{equation}
where $z(t)$ is additive white Gaussian noise (AWGN) with mean being
zero and power spectral density being $N_0/2$ W/Hz.

We assume the receiver can get the perfect channel knowledge.
Received data $r(t)$ is first fed into maximal ratio combining (MRC)
SRAKE in SMPIC. After conventional RAKE processing, the output is
sent to hard-decision module. The estimation of $m$-th bit of each
frame at the output of hard-decision module is denoted as
$\widetilde{b}^{(0)}[m]$. This estimated sequence is then spread,
modulated and processed by a very simple multipath interference
regenerator (MIR). In this sub-module, the modulated sequence is
multiplied by the amplitude of selected $J$ paths and delayed by
corresponding time. So the reconstituted $j$-th path signal can be
expressed as
\begin{equation}
\label{regen}
\begin{aligned}
\widetilde{r}_j^{(0)}(t)=
&\sum\limits_{m=0}^{M-1}\sum\limits_{n=0}^{N-1}a_{k_j,l_j}\widetilde{b}^{(0)}[m]c[n]\\
&g_T(t-T_{l_j}-\tau_{k_j,l_j}-nT_c-mT_b)
\end{aligned}
\end{equation}
where $j\in\{1,2,...,J\}$, and $\alpha_{k_j,l_j}$ is the multipath
gain coefficient corresponding to the $j$-th RAKE finger. $T_{l_j}$
and $\tau_{k_j,l_j}$ denote the delays.

In the next iteration, the input signal to the $j$-th finger is
represented as
\begin{equation}
\label{iteration}
\begin{aligned} r_j^{(1)}(t)=r(t)-w\sum\limits_{j'=1,j'\neq j}^J
\widetilde{r}_{j'}^{(0)}(t)
\end{aligned}
\end{equation}
where $w\in[0,1]$ is a constant named as interference rejection
weight, which allows to reduce the impact of possible errors
presented in the estimated multipath interference replicas. Then
$r_j^{(1)}(t)$ is delivered to the MRC SRAKE receiver in the next
iteration. The SRAKE output can be used as either the input of MIR
for the following iterations, or the output of the SMPIC receiver if
the pre-defined iteration time $p$ is achieved. Finally, the output
of the SMPIC is sent to the LE to reduce the remaining ISI.

\section{Performance and Computational Complexity Analysis}

\subsection{Performance Analysis and the Matched Filter Bound}
In this subsection, the effect of multipath components (MPCs) on
conventional SRAKE receivers and the validity of SMPIC are analyzed
first.

The energy of $g_T(t)$ is defined as $E_g $,
\begin{equation}
\label{energy} E_g = \int_{-\infty}^{+\infty}g_T^2(t)dt
\end{equation}
The normalized autocorrelation function of $g_T(t)$ expresses as
\begin{equation}
\label{autocorr} R_g(\triangle t) =\frac{1}{E_g}
\int_{-\infty}^{+\infty}g(t)g(t+\triangle t)dt
\end{equation}

The channel is assumed perfectly known at the receiver. The local
template of the $\widetilde{m}$-th bit in the $j$-th finger of SRAKE
is given by
\begin{equation}
\label{template}
v_j(t)=\sum\limits_{\widetilde{n}=0}^{N-1}\alpha_{k_j,l_j}^*c[\widetilde{n}]g_T(t-T_{l_j}-\tau_{k_j,l_j}-\widetilde{m}T_b-\widetilde{n}T_c)
\end{equation}
where $(.)^*$ denotes complex conjugation. The correlation output of the $j$-th finger is
\begin{equation}
\label{corr_result}
R_j(t)=\int\limits_{\widetilde{m}T_b}^{(\widetilde{m}+1)T_b}r(t)v_j(t)dt
=b[\widetilde{m}](S+I_1+I_2)+I_3+Z
\end{equation}
where $Z$ represents the effect of noise, and
\begin{equation}
\label{signal} S= NE_g|\alpha_{k_j,l_j}|^2
\end{equation}
is the signal component. $I_1$, $I_2$ and $I_3$ are the IPI, ICI and
ISI respectively,
\begin{equation}
\label{IPI} I_1 =
NE_g\sum\limits_{l=0}^{L-1}\sum\limits_{k=0}^{K-1}a_{k,j}a_{k_j,l_j}^*R_g(T_l-T_{l_j}+\tau_{k,j}-\tau_{k_j,l_j})
\end{equation}
where $l\neq l_j$ or $k \neq k_j$,
\begin{equation}
\label{ICI}
\begin{aligned}
I_2=
&E_g\sum\limits_{l=0}^{L-1}\sum\limits_{k=0}^{K-1}\sum\limits_{n=0}^{N-1}\sum\limits_{\widetilde{n}=0}^{N-1}a_{k,j}a_{k_j,l_j}^*c[n]c[\widetilde{n}]\\
&R_g(T_l-T_{l_j}+\tau_{k,j}-\tau_{k_j,l_j}+(n-\widetilde{n})T_c)
\end{aligned}
\end{equation}
where $n\neq \widetilde{n}$,
\begin{equation}
\label{ISI}
\begin{aligned}
I_3=
&E_g\sum\limits_{m=0}^{M-1}\sum\limits_{l=0}^{L-1}\sum\limits_{k=0}^{K-1}\sum\limits_{n=0}^{N-1}\sum\limits_{\widetilde{n}=0}^{N-1}a_{k,j}a_{k_j,l_j}^*b[m]c[n]c[\widetilde{n}]\\
&R_g(T_l-T_{l_j}+\tau_{k,j}-\tau_{k_j,l_j}+(n-\widetilde{n})T_c\\
&+(m-\widetilde{m})T_b)
\end{aligned}
\end{equation}
where $m\neq \widetilde{m}$.

The accuracy of the RAKE output is closely related to the
statistical properties of IPI, ICI, and ISI, which follow an
impulsive distribution \cite{Interference ICC}. The conventional
symbol level equalizer can only combat long ISI at the cost of high
computational complexity, but fails to mitigate IPI and ICI
effectively. When the SF is small, which means that the
autocorrelation property of the spreading code is poor, the
multipath interferences degrade the performance of the RAKE-DFE
receiver dramatically. From (\ref{regen}) and (\ref{iteration}), we
can see that the proposed SMPIC can subtract the $J$-1 strongest
interference components in every finger before the correlation and
combining at each iteration, hence the strongest interferences,
including $I_1$, $I_2$, and $I_3$, in (\ref{corr_result}) can be
mitigated effectively.

To validate the effectiveness of the SMPIC-LE receiver, in the
following simulation part, the performance of SMPIC-LE is compared
with the MFB of DS-UWB systems, which yields the absolute
performance limit for equalization schemes. In order to obtain
expressions for the bit error rate (BER) of the MFB, we define the
signal-to-noise ratio (SNR) as
\begin{equation}
\label{MFB_SNR} \gamma_r = \frac{E_b(r)}{N_0}
\end{equation}
where $E_b(r)$ is the received energy per bit for the $r$th UWB
channel realization. The corresponding BER($\gamma_r$) for BPSK in
one particular channel realization can be written as
\begin{equation}
\label{MFB_BER} BER(\gamma_r) = Q(\sqrt{2\gamma_r})
\end{equation}
where $Q(*)$ stands for $Q$ function. The average BER is obtained
semi-analytically by averaging over $R$ channel realizations
\begin{equation}
\label{MFB_BER_aver} BER_{MFB} =
\frac{1}{R}\sum_{r=1}^RBER(\gamma_r)
\end{equation}
As for the DS-UWB systems employing $J$-finger RAKE receiver, where
$J$ is much smaller than the total number of resolvable multipath
components for complexity reasons, we obtain
\begin{equation}
\label{MFB_Eb} E_b(r) = \sum_{j=1}^J |h(t_j)|^2
\end{equation}
where $t_j$ ($j\in{1,2,...J}$) denotes the positions of the
strongest $J$ multipath components in one channel realization. The
resolution of $t_j$ equals to the sampling rate at the receiver. In
this paper, the derived MFB takes into account the sampling rate at
the receiver and the effect of selective RAKE combining with a
limited number of RAKE fingers. Therefore, it demonstrates an
accurate performance bound for practical receivers in DS-UWB
systems.

\subsection{Computational Complexity Analysis}


In this paper, the computational complexity of SMPIC-LE and
SRAKE-DFE receiver is calculated in terms of multiplications and
divisions per output symbol (MADPOS) \cite{Proakis}.

The proposed SMPIC is comprised of two kinds of basic sub-modules:
one is the MRC SRAKE, and the other is the MIR. When the SF is
small, correlators, the main part of SRAKE, are quite simple. The
MIR can be seen as the inverse procedure of RAKE processing from
Section II. Therefore, the computational complexity of MIR is
comparable with that of RAKE combining. Moreover, the computational
complexity of SMPIC is independent of the magnitude of path delays,
hence it can be kept under a relatively low level in various
transmission scenarios. The computational complexity of SRAKE and
SMPIC is given by
\begin{equation}
\label{SRAKE-Complexity}
\begin{aligned}
C_{SRAKE}=2\times J
\end{aligned}
\end{equation}
\begin{equation}
\label{SMPIC-Complexity}
\begin{aligned}
C_{SMPIC}=2\times (p+1)\times J +p\times3J
\end{aligned}
\end{equation}
where $J$ stands for the RAKE finger number and $p$ denotes the
iteration time in SMPIC.

For equalization, the widely used adaptive Kalman recursive
least-square (K-RLS) algorithm is employed for adjusting tap
coefficients to ensure fast convergence and lower stead-state mean
square error (MSE), and hence a favorable detection performance in
UWB system \cite{RLS_advan}. The adaptive equalizer works in two
stages: in training stage, a training sequence is employed to
initially adjust the tap weights; in decision directed stage, the
decisions at the output of the equalizer are used to continue the
coefficients adaption process. The computational complexity of
equalizers based on K-RLS is approximately given by \cite{Proakis}
\begin{equation}
\label{CRLS-Complexity}
\begin{aligned}
C_{K-RLS}=2.5\times N^2+4.5\times N
\end{aligned}
\end{equation}
where $N$ represents the total tap number in equalizers.


\section{Numerical Results and Discussion}

\subsection{System Parameters}
Monte Carlo simulations have been run to access the performance of
the proposed receiver in high rate DS-UWB systems with low SF. The
spreading code is set as [-1 +1] with SF being an extreme of 2. The
sampling rate is $T_c/4$. At the receiver, the value of interference
rejection weight $w$ in SMPIC is chosen as 0.9 by investigating the
effect of different weights on the system BER performance. The
iteration time $p$ is set as 2, which can guarantee the performance
convergence in most cases through our simulations. The forgetting
factor in K-RLS algorithm is 0.99999. As for equalization, without
notable instructions, the lengths of LE tap $L$, DFE feedforward tap
$FF$ and feedback tap $FB$ are set as 15, 25 and 20, respectively.
The IEEE 802.15.3a CM1 line-of-sight (LOS) and CM4 extreme non-LOS
(NLOS) UWB indoor channel models are considered here. According to
the recommended instructions in \cite{Channel}, the numerical
results are averaged over the best 90 out of 100 channel
realizations.




\subsection{Bit Error Rate Performance}

As a function of $E_b/N_0$ at the input of receivers, the BER
performance of SRAKE-DFE and SMPIC-LE is evaluated and compared with
MFB.

\begin{figure}[t]
\begin{center}
  \includegraphics[width= 9 cm]{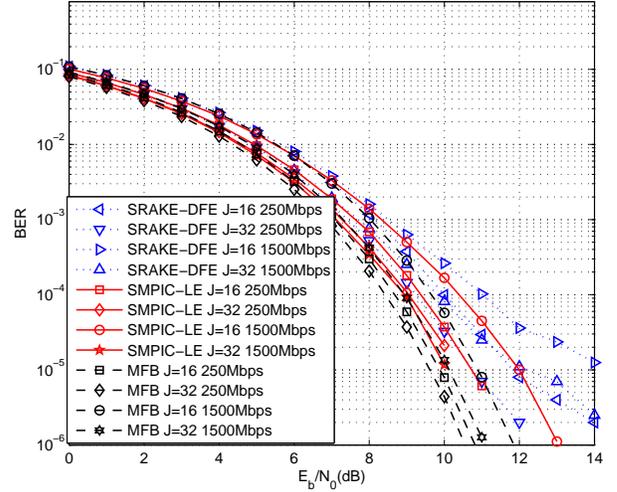}
  \caption{BER performance of SRAKE-DFE and SMPIC-LE receivers in CM1 channel}
  \label{cm1}
\end{center}
\end{figure}

\begin{figure}[t]
\begin{center}
  \includegraphics[width= 9 cm]{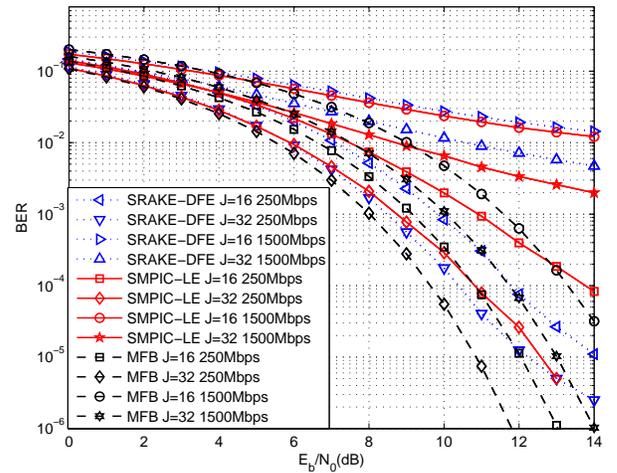}
  \caption{BER performance of SRAKE-DFE and SMPIC-LE receivers in CM4 channel}
  \label{cm4}
\end{center}
\end{figure}



First, we present BER curves of SMPIC-LE and SRAKE-DFE receivers
with different RAKE finger numbers and transmission data rates.
Fig.3 shows the system performance in the CM1 channel model. As seen
in this figure, the SMPIC-LE outperforms the conventional SRAKE-DFE
receiver. When the transmission data rate equals to 250Mbps, the
$J$=32 SMPIC-LE gets a performance gain about 0.2dB over $J=32$
SRAKE-DFE at a BER of $10^{-4}$, and the loss in power efficiency
compared with the derived MFB is within 1dB. As the data rate
increases to 1.5Gbps, the advantage of SMPIC-LE over SRAKE-DFE gets
more significant. It is shown when $J$ equals to 32, the performance
gain is up to about 1dB, and the performance of SMPIC-LE approaches
the MFB well. From Fig.4, it is observed that in the case of CM4
channels, when data rate is 250Mbps, the proposed SMPIC-LE receiver
only suffers negligible performance loss compare with SRAKE-DFE with
the same RAKE fingers. As the data rate increases to 1.5Gbps, the
SMPIC-LE receiver can lower the error floor. From the above two
figures, we can conclude that as the data rate increases, the
performance gain of SMPIC-LE over SRAKE-DFE improves. This can be
attributed to the fact that with the data rate increasing, more
resolvable strong interferences, which degrade the system
performance dramatically, occur at the receiver, and the proposed
SMPIC-LE receiver can alleviate these interferences in a more
effective iterative way compared with the conventional SRAKE-DFE
receiver.

\begin{figure}[t]
\begin{center}
  \includegraphics[width= 9 cm]{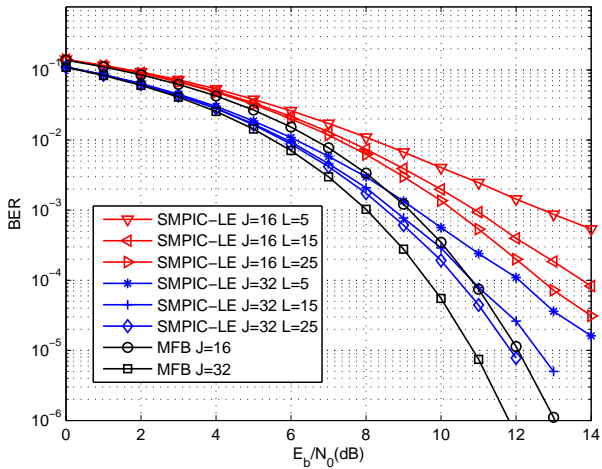}
  \caption{BER performance of SMPIC-LE receivers with different equalizer tap lengths in CM4 channel when the transmission data rate is 250Mbps}
  \label{tap}
\end{center}
\end{figure}

Then the effect of the LE tap number on the BER performance of the
SMPIC-LE receiver is also investigated. From Fig.5, it shows that as
LE tap number $L$ increases, the system performance improves as
well. In addition, as the RAKE finger $J$ increases, the SMPIC-LE
receiver yields a close-to-optimum performance and the performance
gain by increasing LE taps become unobvious. For instance, when the
finger number $J$ is 16, with $L$ increasing from 15 to 25, the
SMPIC-LE receiver can obtain a performance gain of more than 1dB.
When $J$ increases to 32, the performance improvement is only about
0.4dB. This is due to the fact that as the RAKE finger number gets
larger, the strong ISIs are mitigated by SMPIC effectively, hence
increasing LE taps cannot get additional significant performance
gain. This fact provides a useful pointer for system designers when
specifying system parameters. Our findings also suggest that the
SMPIC-LE receiver with more RAKE fingers outperforms the receiver
with less RAKE fingers but more equalizer taps, which demonstrates
the key role of the proposed SMPIC scheme to mitigate severe
multipath interferences in UWB channels.

\subsection{Computational Complexity Comparison}
Finally, the computational complexity of SMPIC-LE and SRAKE-SFE
receivers adopted in the simulations are compared. The MADPOS of
both SMPIC-LE and SRAKE-DFE is shown in Table I. As seen in this
table, when $J$ equals to 16, the MADPOS in the SMPIC-LE receiver
with $L$ = 15 is 822, which is only 15.5\% of that in the SRAKE-DFE
with $FF$ = 25 and $FB$ = 20. When $J$ increases to 32, the SMPIC-LE
can still save 81\% MADPOS than SRAKE-DFE. These results demonstrate
that the computational complexity of SMPIC-LE scheme is much less
than that of conventional SRAKE-DFE receivers.

\newcommand{\tabincell}[2]{\begin{tabular}{@{}#1@{}}#2\end{tabular}}
\begin{table}[!t]
\caption{Computational Complexity Comparison (MADPOS)}
\label{compare} \centering
\begin{tabular}{ccccc}
\hline
&             &  \tabincell{c}{SRAKE-DFE \\($FF$=25, $FB$=20)} &\tabincell{c}{SMPIC-LE \\($L$=15, $p$=2)}  & Saving     \\
\hline
& $J$ = 16 &5297 &822 &84.5\%\\
& $J$ = 32 &5329 &1014 &81.0\% \\
\hline
\end{tabular}
\end{table}



\section{Conclusions}
The scheme presented in this paper offers a low computational
complexity alternative to the conventional SRAKE-DFE receiver, which
provides a more efficient way for UWB signal detection by mitigating
significant multipath interference components specifically. In this
proposed SMPIC-LE scheme, the receiver can alleviate the strongest
IPI, ICI and ISI, while the former two interferences are
unresolvable in conventional RAKE-DFE receivers. The MFB, which
takes into account the effects of sampling rate and the number of
RAKE fingers at the receiver, is also derived. Numerical results and
complexity analysis show that compared with SRAKE-DFE, the SMPIC-LE
receiver, with much lower computational complexity, can achieve
similar or even better performance in high rate DS-UWB systems with
low SF for various UWB propagation scenarios. In addition, as the
RAKE finger number increases, the low-complexity SMPIC-LE receiver
approaches the derived MFB limit well.

\section*{Acknowledgment}
This work is supported by National Nature Science Foundation of
China No. 60928001 and 60972019, National Basic Research Program of
China under grant No. 2007CB310608, and the National Science \&
Technology Major Project under grant No. 2009ZX03006-007-02 and
2009ZX03006-009.


%

\end{document}